# Runtime Optimization of Identification Event in ECG Based Biometric Authentication

Nafis Neehal[1] [*], Dewan Ziaul Karim[1, 2], Sejuti Banik[1], Tasfia Anika[1]

[1] Department of CSE, Daffodil International University, Dhaka-1207, Bangladesh.
[2] Department of CSE, BRAC University, Dhaka-1212, Bangladesh.

nafis.cse@diu.edu.bd, ziaul.karim@bracu.ac.bd,
sejuti.cse@diu.edu.bd, anika.cse@diu.edu.bd

**Abstract.** Biometric Authentication has become a very popular method for different state-of-the-art security architectures. Albeit the ubiquitous acceptance and constant development in trivial biometric authentication methods such as fingerprint, palm-print, retinal scan etc., the possibility of producing highly competitive performance from somewhat less-popular methods still remains. Electrocardiogram (ECG) based biometric authentication is such a method, which, despite its limited appearance in earlier research works, are currently being observed as equivalently high-performing as other trivial popular methods. In this paper, we have proposed a model to optimize the runtime of identification event in ECG based biometric authentication and we have achieved a maximum of 79.26% time reduction with 100% accuracy.

## 1 Introduction

Biometrics is a scientific procedure for person identification and/or verification based on physiological or behavioral characteristics (motoric or cognitive) of individuals. To explain the term "biometrics", the word can be broken down into: bio, as in biological; and metric, as in measurement. So basically biometrics are biological measurements. Currently, biometrics frameworks have become very much incorporated into the fabric of everyday life—deployed where and whenever protected access to a trustworthy instrument is required. Face, fingerprint, retina, iris etc. are some examples of physiological biometrics. Behavioral biometrics includes signature, gait, keystroke etc. But there is another branch of biometrics which has been picking up the thrust over the past decade and that is the use of biological signals (bio signals) such as the electroencephalogram (EEG) or Electrocardiogram (ECG) as biometric characteristics. As a comparatively recent topic in research, entities obtained from ECG signals like Inter Pulse In-terval (IPI) or Heart Rate Variability (HRV) can be efficiently used to identify individuals serving the purpose of a biometric entity. Compared to other biometric systems, ECG based biometric is suitable across a more extensive community of people including amputees. IPI (heart signal) can be obtained from any part of the body (e.g., finger, toe, chest, and wrist). Apart from versatile acquisition, ECG based biometrics contain other facilities such as lower template size, minimal computational requirement, etc. The ECG is basically the graphical representation of the electrical impulses of the heart. Electrical activity of the heart is represented by the ECG signal.

## 2     Background Study

### 2.1    Physiology of ECG

The human heart is made up of four chambers: 2 atria and 2 ventricles (left and right). Blood gets into the heart using superior and inferior vena cava, purging de-oxygenated blood from the body into the right atrium. It is then pumped into the right ventricle and then to the lungs where carbon dioxide is released and oxygen is absorbed. The oxygenated blood then travels back to the left atria, then into the left ventricle from where it is pumped into the aorta and arterial circulation.

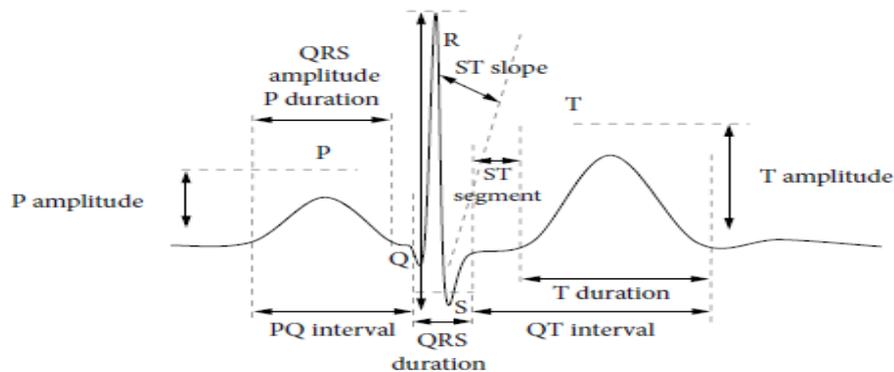

**Fig. 1.** Physiology of ECG Signal and Possible major Fiducial Features [14]

### 2.2    ECG Biometric Usage: Verification & Identification

Such as any other biometric entities, ECG based biometric compares the enrolment ECG against verification ECG or identification ECG. Verification stage approves the claimed identity of a particular person through a Pin or smart card. The person's acquired ECG is matched (one-to-one matching) with his own ECG template, which was procured during an earlier stage of enrolment. On the other hand, during the identification stage, an individual's biometric ECG is recorded and matched throughout the whole ECG template database. After this one-to-many matching, whenever a match is found within a set threshold, the individual is identified.

## 3      Literature Review

FGS Teodoro et al. [1] followed strategies including Genetic Algorithm (GA), Memetic Algorithm (MA) and PSO on the performance of ECG Biometric Systems using KNearest Neighbors, SVM, and a Euclidean Distance Classifier for classification task. MA provided the best result (97.93%). Teodoro et al. worked with more users and produced a more acceptable result than the previous works where accuracy even moved up to 100% (PTB) [2] or 96.44% (Private DB) [3] but with less users.

M. Tantawi et al. [4] conducted a study that quantitatively evaluated the information content of the fiducial based feature set which was subsequently reduced using PCA, LDA, IGR and PASH. PCA indicated that when $t2 = 70$, the FRR is 0 %, yielding an FAR will be 11.7 % [where t1 and t2 are two thresholds]. While, in case of $t2= 85\%$, the FAR was 6.8% and FRR was 7.6%. In LDA, when $t2 = 70$, the FAR was 10.7 % when FRR was 0 %. Later the researcher proposed a feature set named 'PV set' [6] which compared with a super set of 36 fiducial features.

A Lourenço et al. [5] proposed an approach which centered on signals acquired at the subject's hand [32 subjects: 25 males and 7 females with an average age of $31.1\pm9.46$ years.]. In case of Nearest Neighbor (NN) approach, a mean EER of $2.75\%\pm0.29$ and a mean identification error of $5.61\%\pm0.94$ were achieved. In SVM, FAR was 0% and FRR was $13.91\%\pm4:55$. In both cases, an average of 5 heartbeat waveforms were used.

A Page et al. [7] proposed a deep, robust neural network while maintaining a low area and power footprint [memory under 1 MB]. The minimum and maximum computational latency to process a segment took 17.1 and 97.3 ms, respectively. The system was able to attain 99.54% accuracy for QRS complex identification when tested on 90 individuals. It also achieved on average, 99.85% sensitivity, 99.96% specificity, and 0.0582% EER for user identification.

J Sriram et al. [8] proposed a novel ECG and accelerometer-based system where subjects were asked to exercise on the treadmill for 12–15 minutes (training dataset DT) or 5–7 minutes (Test dataset DX). Sitting (DS) and recovery (DR) data were also collected. In total, 10000 samples were taken into consideration. Their protocol forwarded chunks of 4000 samples so that data is sent to the authentication server every 40s.

M. Abo-Zahhad et al. [9] proposed a fusion approach of ECG and PCG for future implementation based on Euclidian Distance and Gaussian mixture models. A later work based on standard 12-lead ECG samples from 20 subjects [10] produced 100% accuracy. Biometric authentication by presenting a set of 15 temporal features [11] provided

100% authentication for 29 subjects. In 2002, researchers in [12] used template matching technique and decision based neural network which provided 100% accuracy in a combined fashion. Researchers in [13] produced an accuracy of 100% for 13 subjects.

In 2010 F. Sufi et al. proposed a novel approach for a self-sufficient system level framework with a new knowledge base in ECG based authentication [14]. The recognition data was never completely same as the enrolment data. To resolve this issue, a threshold is used. The calculations involve PRD, CC, WDM and CL.

After their previous research, F. Sufi introduced Polynomial Distance Measurement (PDM) in ECG based authentication to resolve the open issue of large feature set, random abnormality of ECG etc. [15]. 13 out of 15 subjects displayed 85% of recognition rate remaining within 95% confidence level. PDM misclassified two persons out of a total of 15 subjects, required 340 bytes of template size.

## 4    Proposed Methodology

As stated earlier, the main purpose of this work is to reduce the duration of time taken for identification purpose. Our proposed methodology achieves a maximum of 79.26% time reduction in case of optimal clustering along with 100% accuracy.
Our proposed method has been divided into following phases –

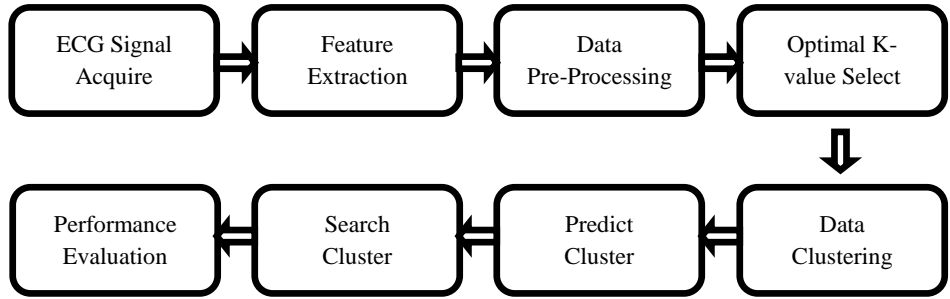

**Fig. 2.** Workflow of proposed model

### 4.1    Dataset Collection

Although our proposed model describes ECG Signal acquiring and Feature Extraction as a requirement, as our primary approach, we have used an existing dataset instead of building one. We have used ECG-ViEW II dataset for our purpose. It has 9 fiducial features, namely – RR Interval, PR Interval, QRS Duration, QT Interval, QTc Interval, P Axis, QRS Axis, T Axis and Age Adjusted Charlson Comorbidity Index (ACCI). The dataset was collected over a 19 year study period (1994-2013) and contains almost 1 million electrocardiograms. This is basically the largest dataset we could find and we needed this huge a dataset because of testing our hypothesis about time reduction by applying clustering in data. We preliminarily kept our scope limited within 50k ECG data. This small version of the dataset was also released by the same community and we used that for our testing purpose.

### 4.2 Data Preprocessing

After the dataset was collected and loaded to our program, at the very beginning, the missing values were filled with zeros. For users having multiple ECG enrolled, the mean of their enrollment values were taken and only the mean value was kept while other entries were removed. After that, floating point data were rounded to nearest integer value for the convenience of calculation. Finally data was transformed scaled (zero mean normalization) and transformed.

Data was rescaled in such a way that they had the properties of a standard normal distribution with $\mu = 0$ and $\sigma = 1$ where $\mu$ is the mean and $\sigma$ is the standard deviation from the mean. Standard scores are calculated as follows –

$$z = \frac{x - \mu}{\sigma}$$

Standardizing the features so that they are centered on 0 with a standard deviation of 1 is not only important if we are comparing measurements that have different units, but it is also a general requirement for many machine learning algorithms.

After this stage, we had roughly 23k entries to work with as some of the entries has been fused with other entries of the same person and the mean of those entries are being considered.

### 4.3 Optimal K Value Select

In this stage we have combinedly used two methods to determine the optimal K value for clustering –

**Elbow Method:**
First of all, Elbow method is a visual method. While applying elbow method, we considered K value ranging [2, 10). Then we calculated SSQ (Sum of squared error) value for each value of K and plotted in a graph.
If in a model, there is a single explanatory variable, then SSQ of that model is given by

$$\text{SSQ} = \sum_{i=0}^{n} \varepsilon^2 = \sum_{i=0}^{n}(y_i - (\alpha + \beta x_i))^2$$

The graph we got after plotting SSQ values against their corresponding K values, we got a graph like this -

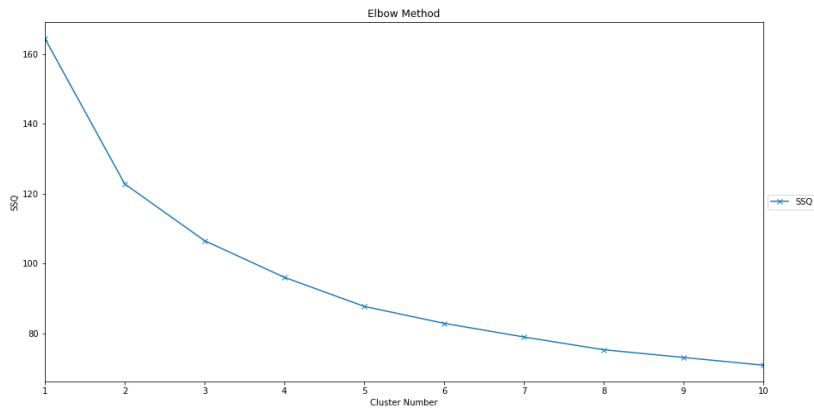

**Fig. 3.** Elbow Method (K vs SSQ)

As we can clearly see, at K=5, the K vs SSQ Curve starts to enter in a plateau from this point as the rate of change or derivative is significantly low and it continues to behave in such way afterwards. So, K = 5 basically indicates the starting point of our plateau which also defines our optimal value K being 5.

**Silhouette Score Maximization:**
S.A. is a way to measure how close each point in a cluster is to the points in its neighboring clusters. It's a neat way to find out the optimum value for k during k-means clustering. Silhouette values lies in the range of [-1, 1]. A value of +1 indicates that the sample is far away from its neighboring cluster and very close to the cluster it's assigned. Similarly, value of -1 indicates that the point is close to its neighboring cluster than to the cluster it's assigned. And, a value of 0 means it's at the boundary of the distance between the two clusters. Value of +1 is idea and -1 is least preferred. Hence, higher the value better is the cluster configuration.

If we now plot average silhouette score value against the K value, we get a graph like Figure 4.

The rule of thumb is to choose the K value with the highest average silhouette score. From this graph we can see that K = 2 is the best, K = 4 is the second best and K = 5 is the third best. We did not choose K = 2 right away as choosing K for our scenario will not solely depend on the cluster quality and optimal partitioning. Rather, we also have to take the overall accuracy and time reduction – these two factors too into our account in order to taking the final decision.

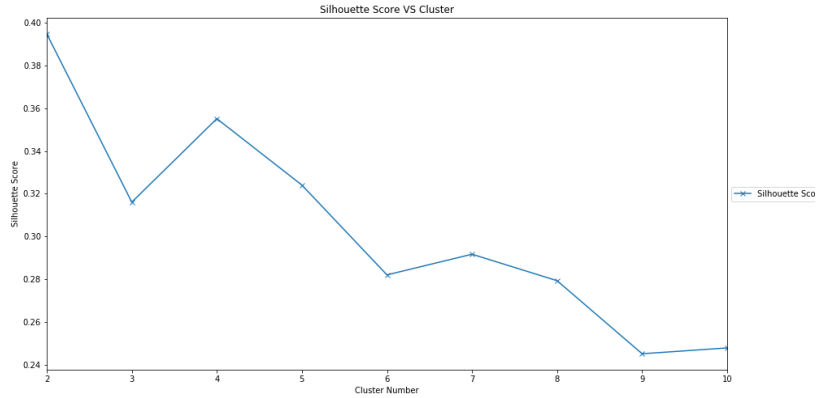

**Fig. 4.** K vs Silhouette Score

But for the time being, our optimal K candidates are 2, 4 and 5. We denote average silhouette score for each K as $s_{avg, k_i}$

### 4.4 Data Partitioning

For testing purpose, instead of choosing only the K candidates (2, 4, 5), we continued to work with all the initial values of K [2, 10). We sequentially clustered the whole dataset into K (2, 3… 9) clusters, created physical partitions by generating separate files and saved it.

The following two pseudocodes are used for data partitioning and time reduction calculation respectively. In a nutshell, suppose, if K = 2, then the whole dataset is split into two new files one for cluster 0 and another for cluster 1. These two files are saved in the same directory. Similarly, if K = 3, then the whole dataset is split into 3 files, namely cluster 0, cluster 1 and cluster 2 and all these three files are saved in a separate directory. In this way partitioning was done for K value up to 9.

*ALGORITHM 1 (dataset): return partitionedDataset*
```
cluster_range := [2, 10)
for each_cluster in cluster_range:
    centroids = each_cluster.Centroids_
    for each_entry in dataset:
        min_cluster_label = min (Eudistance (each_entry, $centroids_{1…n}$))
        assign (each_entry, min_cluster_label)
    sort (dataset, cluster_label, ASC = 1)
    create_physical_partition_according_to_cluster_label()
    return partitioned_files()
```

## 5  Performance Evaluation

We did performance evaluation of our work in several steps –

### 5.1  Time Reduction Calculation

After the partition, the time reduction was calculated. First, a batch of test data was selected. Then, for batch test data X, each of the data point of test data denoted as $x_i$, it's cluster was predicted. Suppose, the prediction returned that this new data point $x_i$ belongs to cluster 2. Then, only cluster 2 file of the current K value was searched. Let us denote time for searching the cluster and find the data as $t_c$. Then, the whole dataset was sequentially searched from for the new data point the time taken is denoted as $t_s$. Then, calculation of time reduction for a single data point would then become –

$$\frac{t_{sx_i} - t_{cx_i}}{t_{sx_i}} \text{ x } 100 \text{ \%}$$

And, then, from this, we calculated the average time reduction for a specific K value, which can be denoted as

$$t_{avg,\, k_j} = \frac{1}{n} \sum_{i=1}^{n} \frac{t_{sx_i} - t_{cx_i}}{t_{sx_i}} \text{ x } 100 \text{ \%}$$

And we calculated this for all $k_j$ (j = 2, 3, … 9) following is the pseudocode –

***ALGORITHM 2 (partitionedDataset, Dataset, testData, K): return timeReduction***
```
filesPartitioned = K
for each_entry in testData:
   min_cluster_label = min(Eudistance(each_entry, centroids₁…ₙ))
   timer1.start()
   access_clusterDataFile(partitionedDataset)
   min_id_c = min(Eudistance(each_entry, entryInClusterFile₁…ₘ))
   timer1.stop()
   timer2.start()
   access_serialDataFile(Dataset)
   min_id_s = min(Eudistance(each_entry, entryInSerialFile₁…ₚ))
   timer2.stop()
return (abs(timer2.time() – timer1.time()) / timer2.time())*100
```

### 5.2  Similarity Measurement

For similarity measurement, we choose two very popular statistical approach for measuring similarity between two vectors, namely – PRD (Percentage Root-Mean-Square Difference) and CC (Cross Correlation).

These two quantities, among which one is the new data point x(i), another one is any entry in the cluster / dataset f(i) are calculated in the following manner –

$$PRD = \sqrt{\frac{\sum_{i=1}^{n}[x(i)-f(i)]^2}{\sum_{i=1}^{n}[x(i)]^2}} \times 100$$

$$CC = \sqrt{\frac{\sum_{i=1}^{n} x(i) \cdot f(i)}{\sum_{i=1}^{n} x(i)^2 \cdot \sum_{i=1}^{n} f(i)^2}}$$

Then we choose 0.5 as weight value for both PRD and CC. Then we calculated the confidence value as such =

$$C = 0.5 * (100 - PRD) + 0.5 * CC$$

We thus calculated confidence values for data point x(i) with all f(i) in the dataset / cluster. For the particular x(i), f(i) pair, for which PRD <=14 and CC >= 0.995 and Confidence value was the maximum, we considered that as a match and returned it as a hit id.

### 5.3   Accuracy Calculation

We take a batch of test data points x(i), and we compared each new data point with all the data points within the selected cluster. If the ID that algorithm returned matched with x(i)'s actual ID, then selected it as a match, otherwise a mismatch.
In such way, we calculated accuracy for each of x(i) from the test set, and then calculated the average accuracy $a_{avg,k_i}$.

### 5.4   Final Decision Logic

In order to finally choose the optimal K value, we have to take into account all the three quantities - $a_{avg,k_i}$, $s_{avg,k_i}$ and $t_{avg,k_i}$
And the decision logic we used for this is a simple weighted sum =

$$D = \underset{i}{maximize} \sum_{i=2}^{9}(w1 * t_{avg,k_i} + w2 * a_{avg,k_i} + w3 * s_{avg,k_i}))$$

Where w1 = 0.2, w2 = 0.5 and w3 = 0.3 was chosen as optimal weight values via trial and error.

## 6  Result Discussion

Our detailed experimental result is given below in Table 1 -

**Table 1. Decision Logic Values**

| K | $x_1$ Time Reduction | $w_1$ 20% | $x_2$ Accuracy | $w_2$ 50% | $x_3$ Silhouette Score | $w_3$ 30% | $\sum_{i=1}^{3} w_i x_i$ |
|---|---|---|---|---|---|---|---|
| 2 | 18.57 | 0.2 | 97  | 0.5 | 0.39 | 0.3 | 52.331 |
| 3 | 57.37 | 0.2 | 100 | 0.5 | 0.32 | 0.3 | 61.57  |
| 4 | 73.10 | 0.2 | 96  | 0.5 | 0.35 | 0.3 | 62.725 |
| 5 | 79.26 | 0.2 | 100 | 0.5 | 0.32 | 0.3 | 65.95  |
| 6 | 80.95 | 0.2 | 94  | 0.5 | 0.28 | 0.3 | 63.27  |
| 7 | 83.43 | 0.2 | 98  | 0.5 | 0.29 | 0.3 | 65.77  |
| 8 | 82.34 | 0.2 | 98  | 0.5 | 0.27 | 0.3 | 65.54  |
| 9 | 86.14 | 0.2 | 97  | 0.5 | 0.24 | 0.3 | 65.8   |

From this chart, we can reach to the conclusion that K = 5 is giving the highest weighted sum value, thus it will be our optimum cluster value. This means, if we divide our whole dataset into 5 clusters, then we can reach to the highest accuracy (100%) with optimal value for silhouette clustering and a max of 79.26% time reduction.

## 7  Future Work

As a part of future work, we would like to include PCA (principal component analysis) technique for our work which might lead to more time reduction. Also, we would like to implement our proposed methodology on the full ECG dataset of 1 Million data.

## 8  Conclusion

If optimized, ECG based authentication can become a very effective and efficient technique for biometric authentication. In order to compete with fingerprint and iris and face detection based authentication it still needs to go a long way. But, hopefully in near future, if optimized properly, it will become the most prominent technique for biometric authentication.